\documentclass[aps,preprint,showpacs,preprintnumbers,amsmath,amssymb]{revtex4-1}
\usepackage{graphicx}
\usepackage{float}
\usepackage[caption=false]{subfig}

\begin{document}

\title{Radiation of inertial scalar particles in the de Sitter universe}

\author{Robert Blaga}
 \email{robert.blaga90@e-uvt.ro}
\affiliation{\small \it
 West University of Timi\c soara,\\
V.  P\^ arvan Ave.  4, RO-300223 Timi\c soara, Romania
}
\begin{abstract}
  We investigate the radiation of an inertial scalar particle evolving in a de Sitter expanding Universe. In the context of scalar QED the process is generated by the first order term in the perturbation theory expansion of the S-matrix. The partial transition probability is obtained and analysed, and soft-photon emission is found to dominate overall. It has been argued that an inertial particle evolving in dS spacetime loses physical momentum just as a decelerated particle in Minkowski space does. It is thus expected that an inertial charge will radiate in a similar way. We investigate the radiated energy and make a qualitative comparison of the angular distribution of the energy with the radiation pattern in the latter case.

\end{abstract}


\maketitle

\section{Introduction}
\qquad De Sitter (dS) spacetime has come to play a very important role in physics,\,both because of its role in early-universe cosmology and because of its implications for the future evolution of our Universe. As a consequence, the literature regarding physics in dS space has bloomed in recent years. \par
A particularly interesting problem which has received some attention in the literature is the quantum radiation from inertial sources. It is well known that in Minkowski space inertial charges do not radiate.\,In dS space the background is dynamic and the constraint of simultaneous energy-momentum conservation is lifted, making otherwise forbidden processes possible.\,Space translation symmetry remains however, giving rise thus to a conserved quantity $p_{cons}$.\,The physical momentum is relatated to this quantity as $p_{phys} = p_{cons} / a$, where $a$ is the scale factor of the spacetime. An inertial particle evolving in dS space loses physical momentum just as an accelerated particle in Minkowski does. It is then expected that it will radiate in a similar way. \par
The classical electromagnetic radiation of an accelerated point-charge in flat spacetime is well known \cite{J1}. The classical problem of inertial and accelerated charges in various charts of dS spacetime has been examined in refs \cite{T1,B1}.\, For comoving observers situated at fixed comoving distance, which have $p_{cons} = 0$, the electric field is purely radial and there is no magnetic field, signaling thus the absence of radiation \cite{A1}.\, Nonetheless, inertial observers with $p \neq 0$ will radiate, with the radiated energy being analogous to that of a linearly accelerated particle in Minkowski space \cite{T1}. The acceleration is proportional to the expansion parameter and to the particle momentum, which can be also understood in terms of geodesic deviation \cite{Mi}. This is in agreement with a number of quantum theoretical results.\par
The groundwork for the investigation of the radiation of charges in a quantum field theoretical framework has been done by Higuchi et al. \cite{H4,H1,H3}. Quantum corrections to the classical radiation (Larmor formula) in flat space have been investigated by Higuchi and Walker \cite{H3}. Nomura, Sasaki and Yamamoto have shown that in the semi-classical (WKB) approximation the radiation emitted by an inertial charged particle in a conformally-flat spacetime takes the form of the classical Larmor formula of an accelerated particle in flat spacetime \cite{N1}.\,The radiation in a time-dependent electric field and in an electromagnetic plane-wave background have been investigated by Nakamura and Yamamoto \cite{N3,N4}. The full quantum radiation in different types of expanding Universes has been analysed in refs \cite{N1,N2}. We mention also that photon emission in a radiation-dominated Universe was investigated in refs \cite{T2,L1}. \par
In the present paper we study the radiation of an inertial scalar particle in the expanding Poincar\'e patch of dS space, in the context of scalar QED, using perturbation theory.  We calculate the transition amplitude from an in-state containing one scalar particle to an out-state containing a scalar particle and a photon. We expect deviations from the classical result due to quantum contributions, but in the region where the WBK approximation is valid (small expansion parameter) we expect that the radiation will coincide with the classical Larmor radiation. \par
We show that the amplitude for photon emission depends only on the quantity $\mu=m/\omega$, with $m$ being the particle mass and $\omega$ the Hubble parameter. It is expected that the effect of the gravitational field will become large when the $\mu$ parameter is of order $\leq 1$. Physically this is accomplished either when (for arbitrary, fixed mass) the expansion factor is large or when the particle Compton-wavelength is comparable to the (arbitrary, fixed) Hubble radius. This latter case may also present some interest even with the present day expansion, because ultra-light particles have received considerable attention in recent years in the literature, mostly in the context of dark matter and dark energy (see for example ref \cite{BB}). \par
The paper is organized as follows. The second section contains the details for the theory of the Maxwell and scalar fields on dS space. In the third section we obtain the transition amplitude and probability, and investigate different asymptotic limits of the parameters, including the flat space limit. In the fourth section we investigate the radiated energy and compare the results to the case of an accelerated charge in Minkowski spacetime. Throughout this paper we work in a system of natural units, in which $\hbar = c = 1$. \\
\section{Preliminaries}
In the present paper we study the process of photon emission by a scalar particle in the expanding Poincar\'e patch of dS space, described by the metric:
\begin{equation}
ds^2 = dt^2 - e^{2wt} d\vec{x}\,^2.
\end{equation}
The Klein-Gordon equation on this background takes the following form:
\begin{equation}
(\partial^2_t - e^{-2\omega t} \Delta + 3\omega\partial_t + m^2)\, \varphi (x,t) = 0
\end{equation}
We consider plane-waves with definite momentum $f_{\vec{p}}(x,t) = f_p(t)\, e^{i\vec{p}\vec{x}}$, the complete set of solutions being \cite{NA,CC1}:
\begin{equation}
\label{H}
f_{\vec{p}}(x,t) = \frac{1}{2}\sqrt{\frac{\pi}{\omega}} \frac{1}{(2\pi)^{3/2}}\, e^{-3\omega t/2}\, e^{i\pi\nu/2} H^{(1)}_{\nu} \left(\frac{p}{\omega}e^{-\omega t}\right)\, e^{i\vec{p}\vec{x}}\, ,
\end{equation}
where we have used the notation $\nu = i\sqrt{\left(\frac{m}{w}\right)^2 - \left(\frac{9}{4}\right)^2}$. \\
The solutions have been chosen such that they select for the vacuum of the scalar field the Bunch-Davies vacuum, fixed by the condition that the modes are of positive frequency in the infinite past (with respect to $i\partial_{t_c}$). We define the notion of particle with respect to the in-modes.  \\
The Maxwell field is comformally invariant, thus we can obtain the solutions to the field equations from the flat-space solutions \cite{CC3}(marked with tilde):
\begin{eqnarray}
A^\mu (x,t_c) &=& \Omega^{-2}(x,t_c)\, \tilde{A}^\mu (x,t_c) \\
W^{\,i}_{\lambda,\vec{k}}(x,t) &=& e^{-2\omega t} \, \frac{1}{(2\pi)^{3/2}}\frac{1}{\sqrt{2k}}\, e^{-ikt_c + i\vec{k}\vec{x}}\, \epsilon^{\,i}{_\lambda} (\vec{k})
\end{eqnarray}
where $\omega t_c = -\,e^{-\omega t}$ is the conformal time, defined so that the metric (1) becomes (explicitly) conformal to Minkowski, with conformal factor $\Omega = e^{\omega t}$. In the second formula we have denoted with $W^{\,i}_\lambda$ the photon wave-function ($i^{th}$ component, with polarization $\lambda$). \\
We follow a general prescription for the quantum theory of interacting fields \cite{BD}, and work with the electromagnetic field in the Coulomb gauge.  \cite{CC2}
\section{Amplitude and Probability of photon emission}
\begin{figure}
\subfloat[\,]{\label{2a}\includegraphics[width=.495\linewidth]{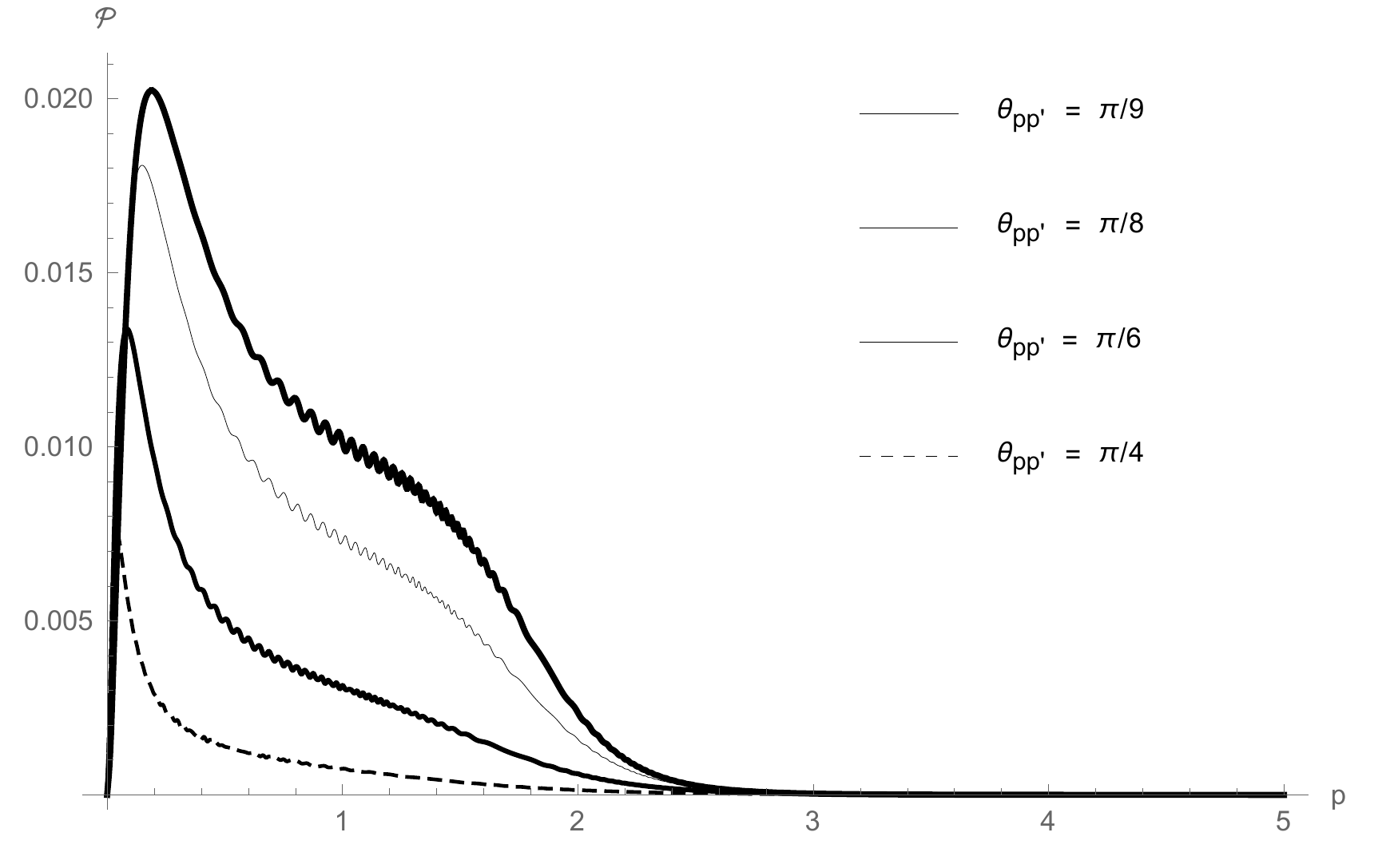}}
\subfloat[\,]{\label{2b}\includegraphics[width=.495\linewidth]{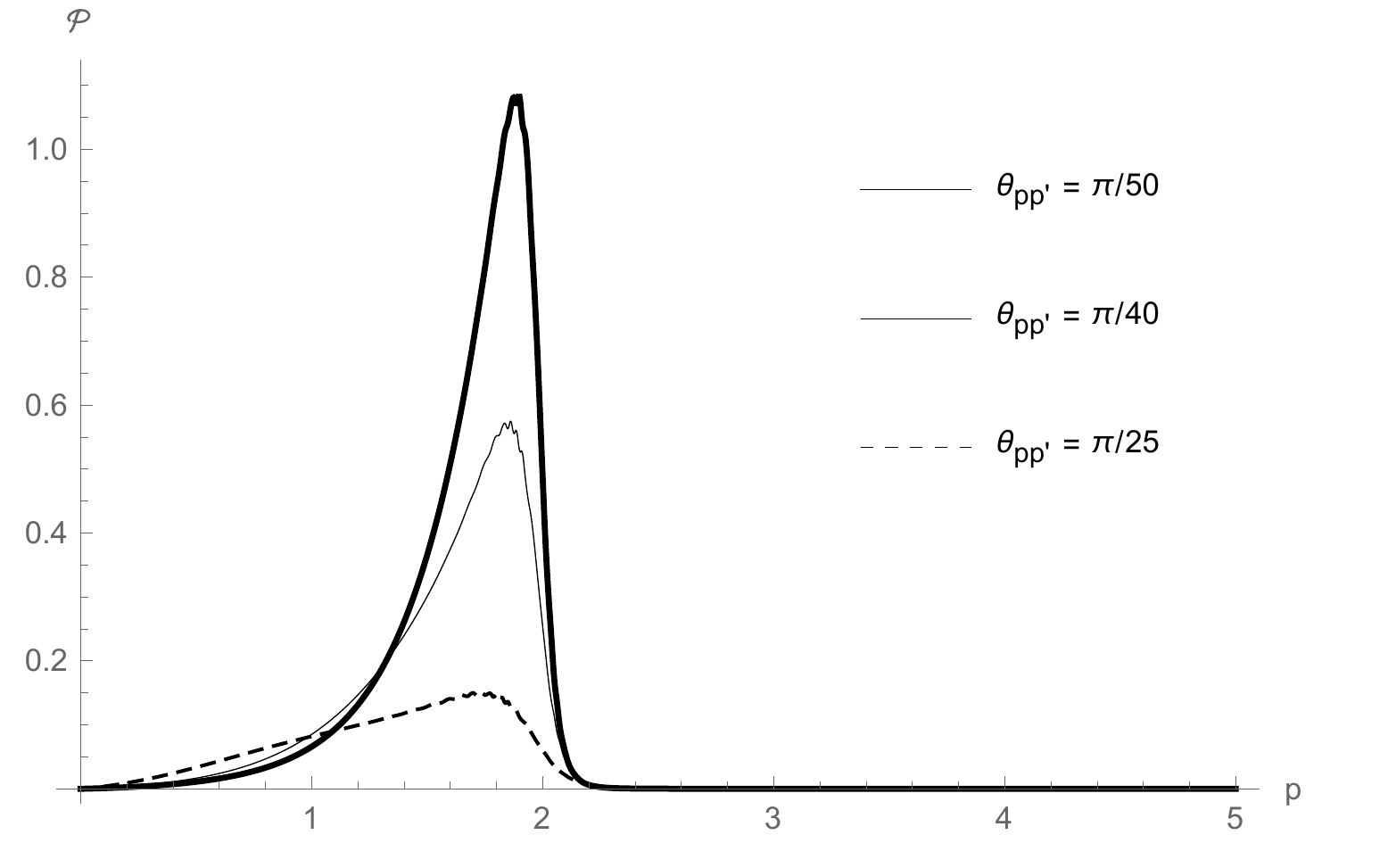}}
\caption{\small \emph{The probability as function of the final scalar momentum. For small angles (right) soft-photon emission dominates and the scalar momentum changes only slightly ($p'\simeq p,\, k \simeq 0$). As we increase the angle (left), the probability for soft-photon drops and another peak appears: it represents the case when the scalar particle transfers most of its momentum to the photon ($p' \simeq k, p \simeq 0$). }}
\end{figure}

We study the transition amplitude from an initial state containing a scalar particle to a final state containing a scalar particle and a photon. More explicitly we are interested in the tree level process, which is just the contribution to the transition amplitude generated by the first order term in the expansion of the S-matrix:
\begin{equation}
S^{\,(1)} = - e \int d^4x\, \sqrt{-g}\, (\varphi^{+}(x) \partial_\mu \varphi(x) - \partial_\mu \varphi^{+}(x)\, \varphi(x)) A^\mu(x).
\end{equation}
The electromagnetic field having only spatial components (Coulomb gauge), will select the spatial derivatives under the integral. As the only space dependent part of the wavefunctions are the momentum plane-waves, the bilateral derivative will draw two factors of momentum and then the spatial integral will result in a Dirac delta function that reinforces momentum conservation. The transition amplitude has the following form:
\begin{eqnarray}
\mathcal{A}_\lambda (\vec{p}^{\, '}, \vec{p},\, \vec{k}) &=& \langle\, 1\, (\vec{p\,})\, , 1\, (\vec{k},\lambda)\, \vert\ S^{\,(1)}\ \vert \, 1\, (\vec{p}^{\, '})\, \rangle\ \ \ \ \qquad \\
&=& -e\int d^4x\, \sqrt{-g}\,\left(f^*_{\vec{p}^{\,'}}(x) \partial_i f_{\vec{p}}(x) - \partial_i f^*_{\vec{p}^{\,'}}(x)f_{\vec{p}}(x)\right)\, W^i_{\lambda,\vec{k}}(x,t) \nonumber\\
 &=&-\,\delta^{(3)}( \vec{p}^{\,'} -\, \vec{p} - \vec{k})\frac{e\pi(\vec{p}^{\, '}+\,\vec{p}\,)\cdot \vec{\epsilon}(\vec{k})}{(2\pi)^{3/2}\sqrt{32k}}\int\limits^\infty_0d\eta\ \eta\,  H^{(1)}_\nu(p'\eta)H^{(2)}_\nu(p\,\eta)\,e^{-ik\eta} \nonumber
\end{eqnarray}
where we have changed the integration variable to $\omega\eta = e^{-\omega t}=-\omega t_c$, and the integration measure is $\sqrt{-g} = e^{3\omega t}$. In order to do the integral we need to express the Hankel functions in terms of Bessel functions of the first kind,
\begin{equation}
H^{(1,2)}_\nu(z) = \pm\ \frac{J_{-\nu}(z) - e^{\mp i\pi\nu} J_\nu (z)}{i\,\sin(\nu\pi)}
\end{equation}
resulting in two types of integrals which can be done analytically:
\begin{equation}
\label{E9}
\int^\infty_0 d\eta\ \eta\,H^{\mathrm{(1)}}_{\,\nu}(p'\eta)\, H^{(2)}_{\,\nu}(p\,\eta)\,e^{-ik\eta} =
\end{equation}
\begin{equation}
=\frac{1}{\ \sin(\pi\nu)^2} \left\{B^+_k(p',p) + B^-_k(p',p)- e^{- i\pi\nu}\, C_k(p',p)- e^{+ i\pi\nu}\, C_k(p,p')\right\} \nonumber\\
\end{equation}
The integrals can be found in refs. \cite{GR,PB},
\begin{eqnarray}
B^\pm_k (p',p) &=& \int^\infty_0 d\eta\ \eta\,J_{\pm\nu} (p'\eta)\, J_{\pm\nu}(p\,\eta)\,e^{-ik\eta - \epsilon\eta}  \\
&=&- \frac{\,ik\ \,}{\pi(p p')^{3/2}}\,\frac{d}{dz} Q_{\pm\nu - \frac{1}{2}} \left(z + i\epsilon\right) \nonumber\\
C_k (p',p) &=& \int^\infty_0 d\eta\ \eta\, J_{\,\nu} (p'\eta)\, J_{-\nu}(p\,\eta)\,e^{-ik\eta - \epsilon\eta}  \\
&=& \left(\frac{p'}{p}\right)^\nu\,\left(\frac{1}{ik}\right)^2\, \frac{\sin(\pi\nu)}{\pi\nu\,} \, F4\left(1,3/2,1+\nu,1-\nu,\frac{\,p^{'2}}{k^2} + i\epsilon, \frac{\,p^2}{k^2}+i\epsilon\right),\nonumber
\end{eqnarray}
where we have denoted $ z = \frac{p^{'2} + p^2 - k^2}{2pp'}$, which is just the cosine of the angle between the initial and final scalar particle momentum if we take into account the momentum conservation.\,We have introduced the factor $\epsilon$ because the argument of the Legendre functions varies between $\left\{-1,1\right\}$, where the function has a branch-cut. It has been chosen so that it reinforces the convergence of the integrals, selecting the correct analytic form for the Legendre functions\cite{CC2}. The vanishing limit should be understood in the results. In the explicit calculations we take $\epsilon$ to be a very small, but finite positive constant.
The result is given here in terms of Legendre functions (Q) and Appell hypergeometric functions (F4). This can be further developed in terms of Gauss hypergeometric functions (F), Gamma functions ($\Gamma$) and Appell hypergeometric functions of the first kind (F1) which can be used for numerical analysis \cite{GR,Sch}. The analytical calculation leading to the final form of the integral can be found in the Appendix. \\
To calculate the total transition probability we need to take the square modulus of the amplitude, average over the helicities of the photon and then integrate over the momentum of the photon and the momentum of the scalar particle. The partial probability is also interesting to study. By integrating over only either the photon or the scalar particle momentum (easily done due to the delta function), we can investigate various limiting cases of the parameters (small/large expansion parameter, small/large momenta) and we can also study the variation with the scalar particle/photon angle. \\
First we deal with the polarization dependent term. If we use the momentum conservation and the fact that the polarization vectors and the wave-vector form an orthogonal basis, we see that this term is just the magnitude of the projection of the initial momentum onto the plane orthogonal to the wave-vector:
\begin{eqnarray}
\frac{1}{2}\sum_\lambda \left\vert (\vec{p}^{\, '} +\, \vec{p}\,)\cdot \vec{\epsilon}(\vec{k}) \right\vert^{\,2} &=& 2\left({p'}^{2} - \frac{(\vec{p'}\cdot \vec{k})^2}{k^2}\right)\\
&=& 2\,{p'}^2 \sin^2\theta_{kp'} \nonumber\\
&=& \frac{2\,{p'}^2 p^2 \sin^2\theta_{pp'}}{k^2}\nonumber
\end{eqnarray}
The above can be easily verified by considering explicit polarization vectors (for example circular polarization). The result does not depend on the orientation of the vectors.\,The $\theta$ variables are the angles between the momenta shown as indices. With this the (partial) transition probability becomes:
\begin{eqnarray}
\mathcal{P}(p',p, \theta_{pp'}) &=& \frac{1}{2}\, \sum_\lambda \, \int d^{3}k\ \left\vert \mathcal{A}_\lambda (\vec{p}',\vec{p},\vec{k}) \right\vert^{\,2} \\
&=& \frac{e^2\pi^2}{16\sin^4(\pi\nu)}\,\frac{{p'}^2 p^2 \sin^2\theta_{p'p}}{(2\pi)^3 \,k^3}\ \nonumber\\
  &\times& \left\vert B^+_k(p',p) + B^-_k(p',p)- e^{- i\pi\nu}\, C_k(p',p)- e^{+ i\pi\nu}\, C_k(p,p')\right\vert^{\,2},\nonumber
\end{eqnarray}
\begin{figure}
\centering
\subfloat[\,]{\label{1a}\includegraphics[width=.495\linewidth]{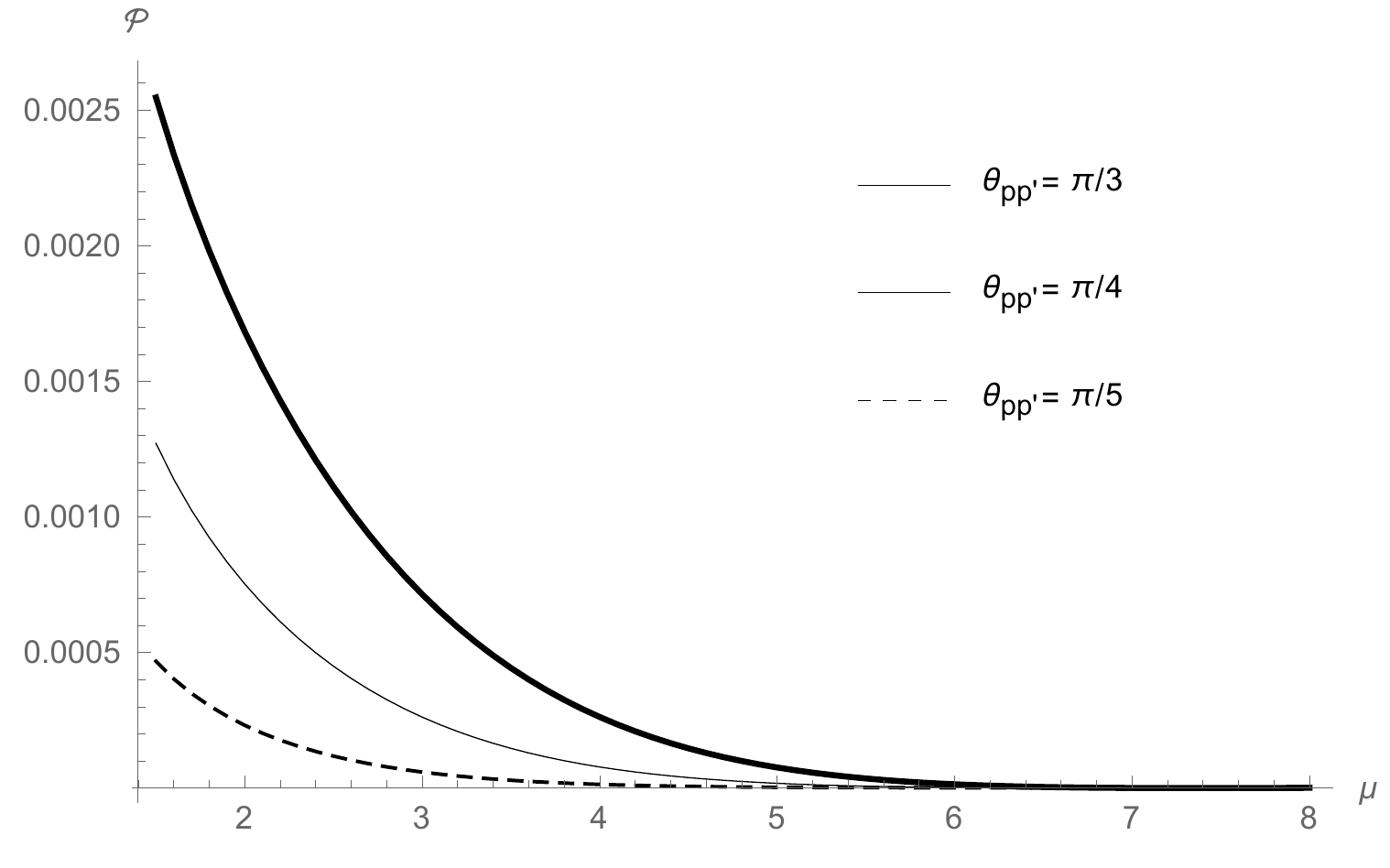}}
\subfloat[\,]{\label{1b}\includegraphics[width=.495\linewidth]{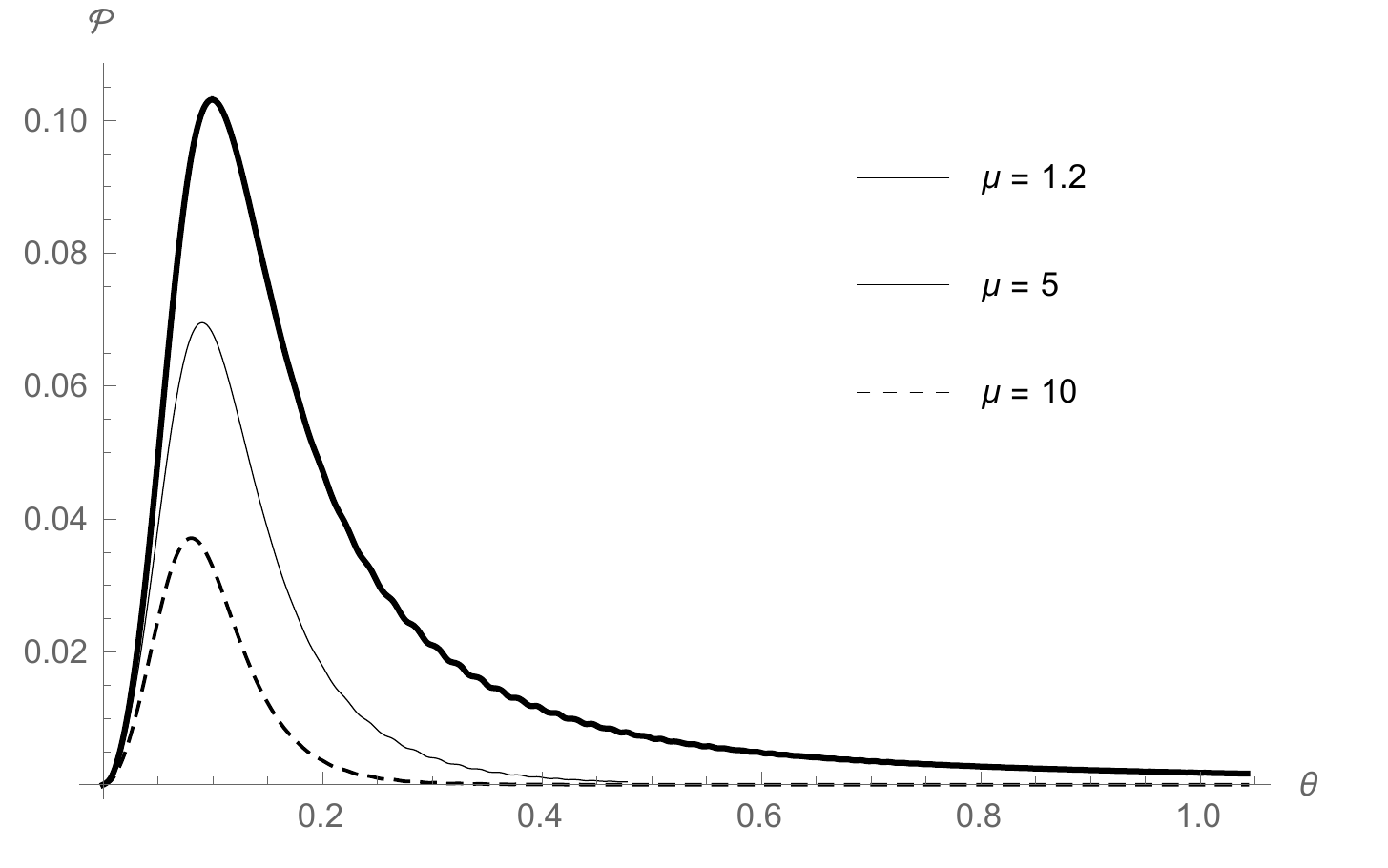}}
\caption{\small \emph{The $\mu$-dependence of the emission probability. Plot\ (a)\  shows how, for fixed momenta
 ($p'=2, p=1$),\,the amplitude declines with the decrease of the expansion parameter, vanishing in the flat limit. Plot (b) shows the probability as a function of the angle between the initial and final scalar momentum ($\,p'=2,p=1\,$), for different values of the expansion parameter.}}
\end{figure}

where it is understood that $ k = \sqrt{ p^{'2} + p^{\, 2} - 2pp'\cos \theta_{pp'}}$. \\
As we have specified earlier, the process is forbidden in Minkowski space because energy-momentum is not conserved. In dS space the background is dynamic and energy can be drawn from and ceded to it, thus making the process possible. We see from fig.[\ref{1a}] that the probability drops rapidly with the decrease of the expansion parameter, vanishing in the flat limit  ($\,\mu = m/\omega \rightarrow \infty $). One could attempt to compute this limit by approximating either the amplitude or the Hankel functions in eq.\,(\ref{H}). \cite{CC4} \\
Graphical analysis of the probability reveals several interesting facts. There are two regimes that have significant probabilities: on one hand there is the emission of small momentum "soft" photons, which do not modify the scalar momentum significantly, and on the other hand there is the process where the scalar particle cedes a considerable amount of its momentum to the photon (we will call these, "hard" photons). \\
In fig.[\ref{1b}] we see the angular distribution of the probability as a function of the angle between the initial and final scalar momenta. The maximum of the curves is at small angles and approaches zero as the expansion parameter decreases. This suggests that soft-photon emission is the most significant contribution in all conditions. This is easy to understand as this process is peaked around $p' \simeq  p + k\,$, which is reminiscent of energy-conservation. As we get closer to flat space the probability overall decreases and soft photon emission becomes dominant. Also note that for values below the threshold of $\mu = 3/2$, nothing spectacular happens as the index of the Hankel functions in eq.\,(\ref{H}) passes from imaginary to real, other than the fact that there is more significant emission at higher angles compared to the weak-gravity case. \\
Figures [\ref{2a}] and [\ref{2b}] show the distribution of the probability as function of the final scalar momentum. It is confirmed that the overall dominant process is soft photon emission at small angles $\theta_{pp'}$, which modifies the scalar momentum only slightly ($p \simeq  p^{'} $). As we look at larger and larger angles the balance tips in favor of hard photon emission.\,For example if we consider the emission leaving the scalar particle momentum at a right angle, the photon has then to compensate for the initial and the final scalar momentum (momentum is conserved on each axis) and thus has to draw a significant amount of energy from the background. This is clearly overall suppressed compared to soft photon emission, but at a fixed large angle the most efficient channel for the process is that in which the final scalar momentum is smallest (this can be seen in fig.[\ref{2a}]).  \\
\section{Radiated Energy}
\begin{figure}
\centering{\includegraphics[width=.55\linewidth]{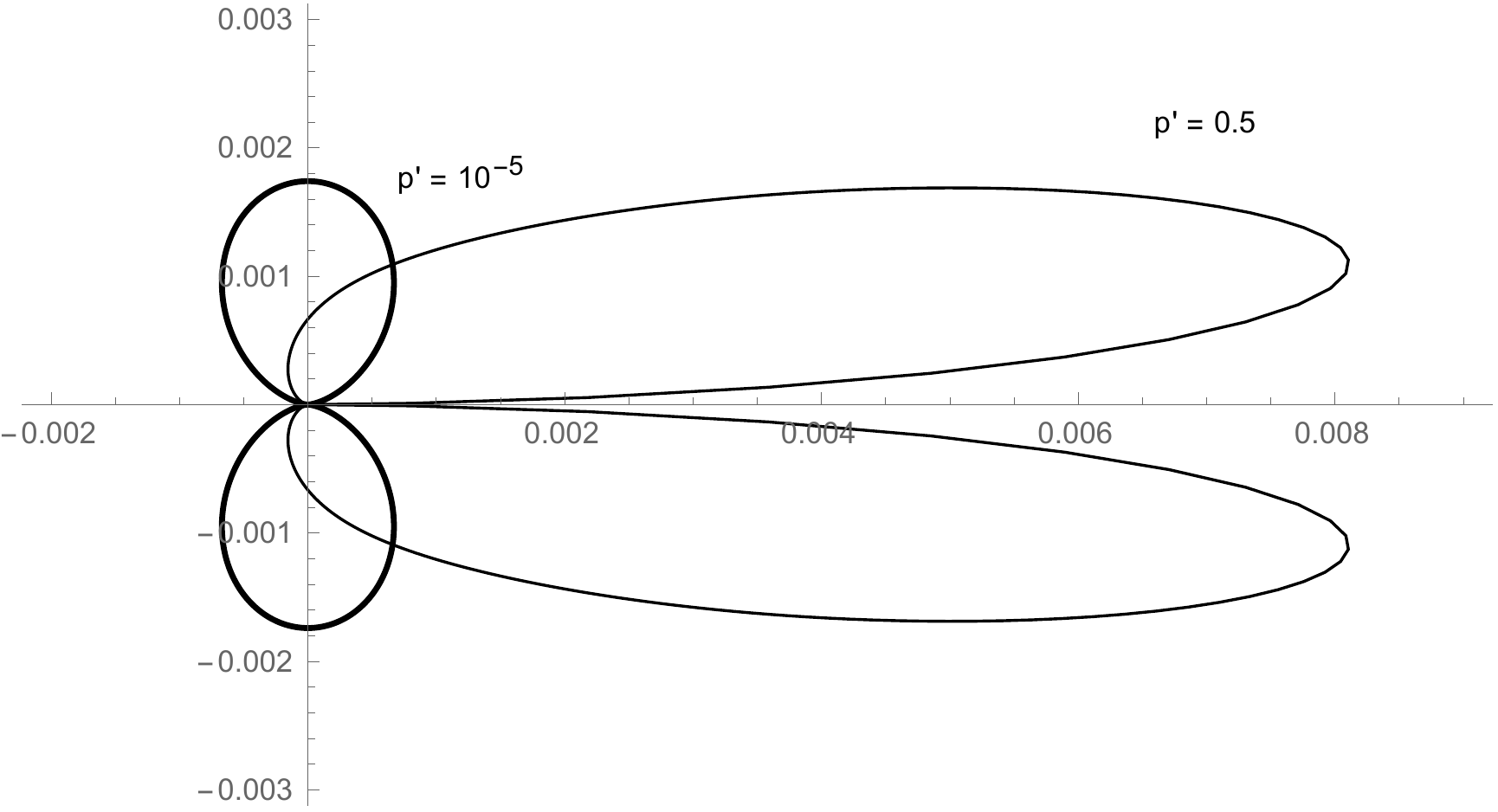}}
\caption{\small \emph{The angular distribution of the radiated energy. We have fixed the scalar mass to unity, thus small momenta means non-relativistic particle and large momentum is the relativistic regime. The graphs where rescaled in order to fit in the same plot (the small momentum graph has been multiplied with $10^5$)}}
\label{4}
\end{figure}

We consider now the radiated energy. This can be calculated from the transition probability, considering that each emission gives a photon in the final state. The total emitted energy is just the sum of the energy of these photons.

\begin{equation}
\mathcal{E} = \frac{1}{2}\, \sum_\lambda \int d^3k\, d^3p\ \,\hbar k \ \left\vert \mathcal A \right\vert^2,
\end{equation}
where the $\hbar$ was restored for clarity. \\
The change in the scalar momentum as a result of photon emission is due to the radiation reaction. We wish to study the angular and frequency distribution of the radiation, irrespective of the backreaction, thus we integrate over the final scalar momentum. As we have mentioned above, the inertial particle in dS space loses momentum similar to an accelerated particle in Minkowski space. We expect thus, that the radiation pattern will be qualitatively similar to that of an accelerated particle in Minkowski (with acceleration parallel to the velocity) \cite{J1}. \\
 \begin{figure}
\subfloat[\,]{\label{3a}\includegraphics[width=.495\linewidth]{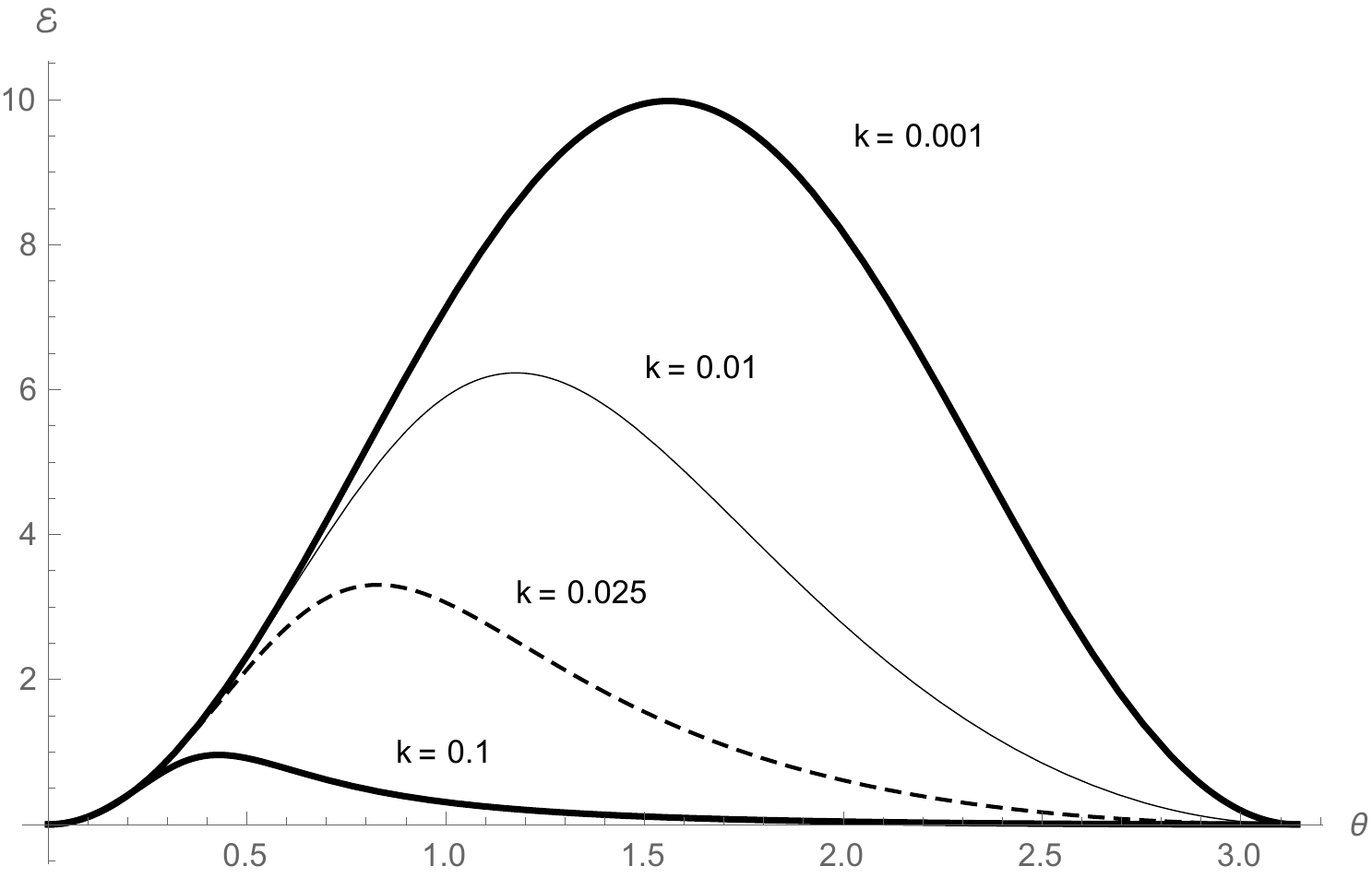}}
\subfloat[\,]{\label{3b}\includegraphics[width=.38\linewidth]{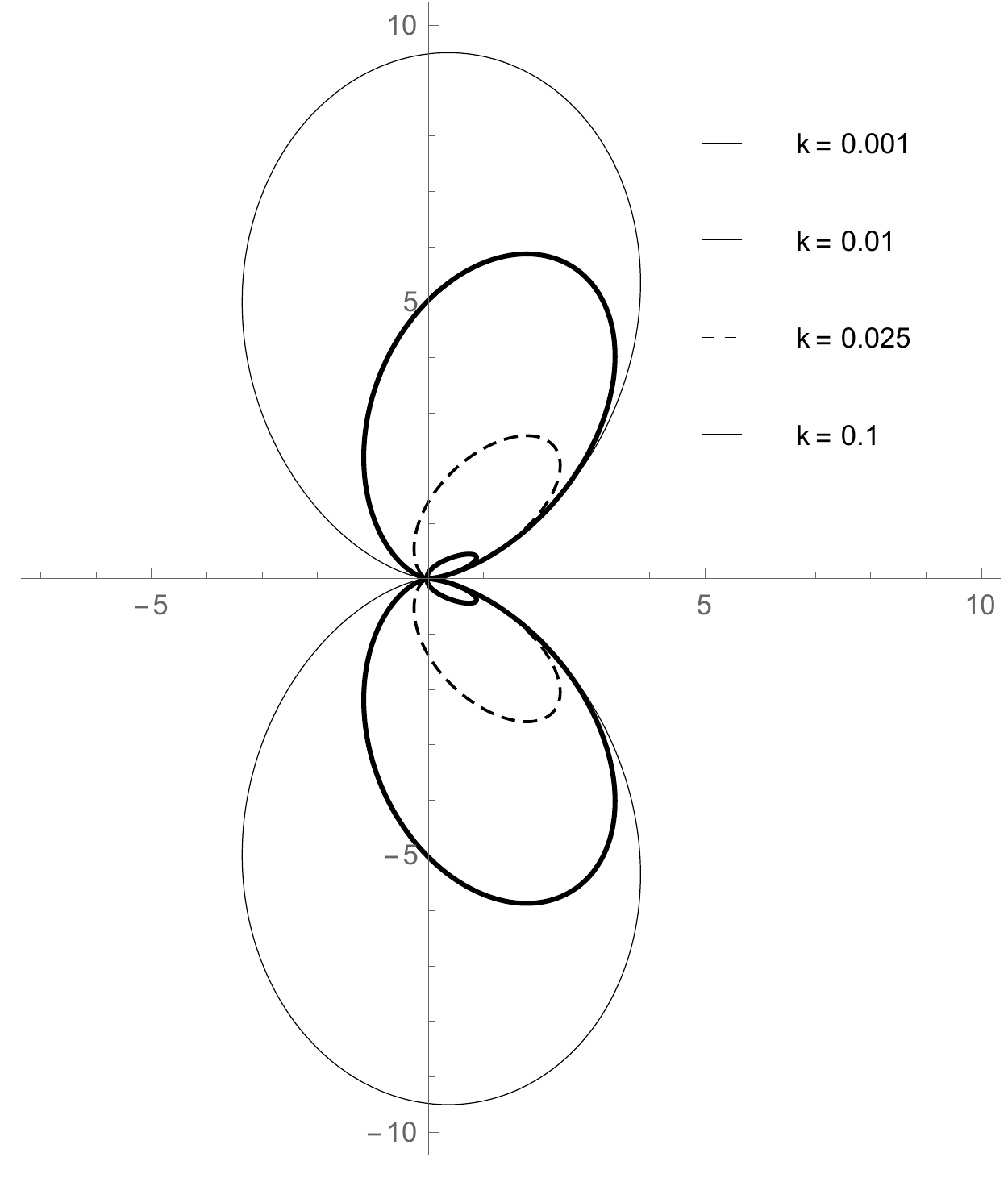}}
\caption{\small \emph{The angular distribution of the (partial) radiated energy ($\mu = \sqrt{2})$, in cartezian (left) and polar (right) coordinates, for different values of the photon momentum.}}
\end{figure}
We consider the case $\mu = \sqrt{2},\, \nu = 1/2$ where the Hankel functions in eq.\,(\ref{H}) have simple analytical forms that allow us to evaluate the amplitude:
\begin{eqnarray}
  \int \limits_0^\infty d\eta\, \eta\ H^{(1)}_\frac{1}{2} (p'\eta)\,H^{(2)}_\frac{1}{2} (p\,\eta)\,e^{-ik\eta-\epsilon\eta}   &=& \frac{2}{\pi}\frac{1}{\sqrt{pp'}}\int \limits_0^\infty d\eta\ e^{i(p'-p-k+i\epsilon)\eta}  \\
 &=& \frac{2}{\pi\sqrt{pp'}}\frac{i}{p'-p-k+i\epsilon} \nonumber
\end{eqnarray}
Note that this is already a considerably strong gravitational field as $ m \sim \omega $. \\
In figure [\ref{3b}] we have plotted in polar coordinates the emitted energy as a function of the emission angle, for different values of the modulus of the photon momentum (the frequency). From the graphs we conclude that soft photons are primarily emitted perpendicular to the motion, while hard photons are mostly emitted at small angles, and that soft photon emission dominates overall. \\
To obtain the angular distribution of the total emitted energy we integrate over the modulus of the photon-momentum. The polar plots for different initial momenta are shown in fig.[\ref{4}]. Remarkably (but expectedly) the behaviour is similar to the Minkowski case: for small initial momentum (for fixed mass, it means small initial velocity) the emission is mainly perpendicular and has the characteristic $\sin^2$\,distribution; if we increase the initial momentum (relativistic regime) the graph tilts towards the x-axis and the radiation is emitted in a narrow cone around the initial direction. The qualitative difference compared to the Minkowski case is that here we have significant emission at higher angles even for a highly relativistic particle.\\
The total energy can be calculated by integrating over all angles. With a small but finite $\epsilon$, as we have considered in our analysis, the total energy is finite but it diverges as we take the vanishing limit. \par The divergence could theoretically stem from either end of the time integral in the amplitude. The contribution of small conformal times $\eta \rightarrow 0$ (future infinity), to the amplitude, is of the form (\ref{E9})
\begin{equation}
\int_{\eta \rightarrow 0} d\eta\ \eta\,\left( A\, \eta^{\,\nu} + B\, \eta^{-\nu} + C\right)
\end{equation}
 where $\nu = -\sqrt{9/4 - \mu^2}$ is the index of the Hankel functions and \emph{A,B,C} are constants with respect to $\eta$. This is divergent for $\mu^2 < 5/4$. A thorough exposition and comprehensive bibliography on this type of (IR) divergence can be found in ref\, \cite{A3}. For this reason, we have only considered the case $\mu^2 > 5/4$.  \\
 For $\eta \rightarrow \infty$ (past infinity) the constribution is of the form
 \begin{equation}
\int^{\eta \rightarrow \infty}d\eta\ \ e^{i(p' - p - k)\eta}
\end{equation}
diverging as the exponent vanishes. Due to the momentum conservation, this is accomplished only when $k \rightarrow 0, p \simeq p'$.
\,This is the type of divergence (often called IR or adiabatic catastrophe) \cite{PK,IZ,A5} that typically arises in theories where massless particles are being radiated.\par
In the case of bremsstrahlung in the field created by a nucleus, altough the number of emitted particles is infinite, the energy is finite in both classical and QED calculations. \cite{IZ}. In the case of a point-charge accelerated in a constant electric field however, the radiated energy diverges.\cite{Ni}
This can be easily seen from the fact that the instantaneously emitted energy (Larmor formula) for a particle with constant acceleration (produced by a constant external field) is finite.\cite{J1} It follows that during the infinite time of the process an infinite amount of energy is being radiated. This is the case also for an inertial classical charge evolving in dS spacetime \cite{T1}, as this process can be interpreted as bremsstrahlung in a constant external gravitational field. \par
The parameter $\epsilon$, that we have introduced in our amplitude, plays the role of a cut-off for infinite conformal time (infinite past in cosmological time) thus rendering the total emitted energy finite. \par
Finally, we note that the quantity under study is the energy in the conformally-flat spacetime. The physical electromagnetic energy, being $E = \mathcal{E}/a$, is redshifted away in the infinite future because of the infinite expansion of space.
\section{Concluding remarks}
We have investigated the process under which a photon is emitted by an inertial scalar particle evolving in the expanding spatially flat FRW chart of dS spacetime. We have studied the transition probability for the process and found that the overall dominant channel is soft photon emission. In the second part of the paper we have analysed the angular distribution of the radiated energy. We have found that it is qualitatively similar to that of an accelerated particle in flat spacetime. By this we mean that for non-relativistic particles the radiated energy is small and mainly directed perpendicular to the motion, while relativistic particles radiate strongly in the forward direction, the radiation being concentrated in a narrow cone around the direction of motion. \\
We will extend the qualitative and quantitative comparison with the flat spacetime case in a future paper. We will first need to find a correct definition of the rate of radiation of energy in our case, then compare it with the Larmor formula \cite{J1}. Finally we will extend the study to fermions.

\appendix
\section{Analytical expression of the amplitude}
In order to process the first integral ($B^\pm$), we note that the argument of the Legendre function $z = \frac{p^{'\,2} + p^2 - k^2}{2pp'} \equiv  \cos(\theta_{pp'}) $ takes values from the interval $\{-1,1\}$, where the Legendre functions have a branch-cut, thus we need to use the following relations \cite{GR}:
\begin{eqnarray}
Q_\xi(z \pm i\epsilon) &=& \frac{\pi}{2\sin\pi\xi}\,\left(\,e^{\mp i\pi\xi} P_\xi(z) - P_\xi(-z)\right) \\
P_\xi(z) &=& F\left(-\xi,1+\xi,1,\frac{1-z}{2}\right) \\
\frac{d}{dz} F(a,b,c,z)&=& \frac{ab}{c}\, F\,(1+a,1+b,1+c,z),
\end{eqnarray}
thus
\begin{eqnarray}
B^\pm_k (p',p) &=& \int^\infty_0 d\eta\ \eta\,J_{\pm\nu} (p'\eta)\, J_{\pm\nu}(p\,\eta)\,e^{-ik\eta - \epsilon\eta}  \\
&=&- \frac{\,ik\ \,}{\pi(p p')^{3/2}}\,\frac{d}{dz} Q_{\pm\nu - \frac{1}{2}} \left(z + i\epsilon\right) \nonumber
\end{eqnarray}
\begin{equation}
= \frac{\,ik\ \,}{(p p')^{3/2}}\,\frac{\left(\nu^2 - \frac{1}{4}\right)}{4\cos\pi\nu}\left[ie^{\mp i\pi\nu}F\left(\frac{3}{2}\mp\nu,\frac{3}{2}\pm\nu,2,\frac{1-z}{2}\right) +F\left(\frac{3}{2}\mp\nu,\frac{3}{2}\pm\nu,2,\frac{1+z}{2}\right) \right] \nonumber
\end{equation}
\\
In order to evaluate the $2^{nd}$ integral, we use the following formula \cite{PB}:
\begin{eqnarray}
\int \limits_{0}^{\infty} dz\,z^{-\lambda}\, J_\mu(p'z)\,J_\nu(p\,z)\,e^{-ikz} &=& \frac{p^{'\mu} p^{\,\nu}}{2^{\mu+\nu}\,(ik)^{\lambda+\mu+\nu+1}}\frac{\Gamma\left(\lambda+\mu+\nu+2\right)}{\Gamma(1+\mu)\Gamma(1+\nu)}
\end{eqnarray}
\begin{equation}
\qquad \qquad \times\,F4\left(\frac{\lambda+\mu+\nu+1}{2},\frac{\lambda+\mu+\nu+2}{2},1+\mu,1+\nu,\frac{p^{'2}}{k^2},\frac{p^{\,2}}{k^2}\right) \nonumber
\end{equation}
The F4 Appell hypergeometric function is absolutely convergent for
\begin{equation}
\left\vert\sqrt\frac{p^{'2}}{k^2}\,\right\vert+\left\vert\sqrt\frac{p^{\,2}}{k^2}\,\right\vert < 1,
\end{equation}
a condition that is not met by our physical momenta, because the momentum conservation limits the photon momentum to the interval $k \in \{\vert \,p\,-p'\vert,p\,+p'\,\}$. We need then to find a suitable analytical continuation \cite{GR,Sch}:
\begin{eqnarray}
F4\left(a,b,c,c',x,y\right) &=& \frac{\Gamma(c')\,\Gamma(b-a)}{\Gamma(c'-a)\,\Gamma(b)}\,(-y)^{-a}\,F4\left(a,a+1-c',c,a+1-b,\frac{x}{y},\frac{1}{y}\right) \nonumber\\
&+& \frac{\Gamma(c')\,\Gamma(a-b)}{\Gamma(c'-b)\,\Gamma(a)}\,(-y)^{\,b}\,\ \,F4\left(b+1-c',b,c,b+1-a,\frac{x}{y},\frac{1}{y}\right) \nonumber
\end{eqnarray}
\begin{equation}
F4(a,b,c,b,x(1-y),y(1-x))=
\end{equation}
\begin{equation}
= (1-x)^{-a}(1-y)^{-a} F1 \left(a,1+a-c,c-b,c,\frac{xy}{(1-x)(1-y)},\frac{x}{x-1}\right) \nonumber
\end{equation}
\begin{equation}
F4\left(a,b,1+a-b,b,\frac{1}{(1-x)(1-y)},\frac{1}{1-y)(1-x)}\right) =
\end{equation}
\begin{equation}
= (1-y)^a\,F\left(a,b,1+a-b,-\frac{x(1-y)}{1-x}\right) \nonumber
\end{equation}
The integral then becomes:
\begin{eqnarray}
C_k (p',p) &=& \int^\infty_0 d\eta\ \eta\, J_{\,\nu} (p'\eta)\, J_{-\nu}(p\,\eta)\,e^{-ik\eta - \epsilon\eta}  \\
&=& \left(\frac{p'}{p}\right)^\nu\,\left(\frac{1}{ik}\right)^2\, \frac{\sin(\pi\nu)}{\pi\nu\,} \, F4\left(1,\frac{3}{2},1+\nu,1-\nu,\frac{\,p^{'2}}{k^2} + i\epsilon, \frac{\,p^2}{k^2}+i\epsilon\right) \nonumber\\
&=&-\left(\frac{p'}{p}\right)^\nu\,\left(\frac{1}{ik}\right)^2\, \frac{\sin(\pi\nu)}{\pi\nu\,}  \nonumber\\
&\times \{& -\frac{k^{\,2}}{p^{\,2}}\frac{2\nu}{(1-x_1)(1-y_1)}\,F1\left(1,\frac{3}{2},-\frac{1}{2}-\nu,\frac{1}{2},\frac{x_1\,y_1}{(1-x_1)(1-y_1)},\frac{x_1}{(x_1-1)}\right) \nonumber \\
&+&2\sqrt{\pi}\frac{\Gamma(1-\nu)}{\Gamma(-\frac{1}{2}-\nu)}\left(-\frac{p^{\,2}}{k^2}\right)^{\frac{3}{2}}(1-y_2)^{\frac{3}{2}+\nu}\,F\left(\frac{3}{2}+\nu,\frac{3}{2},1+\nu,-\frac{x_2(1-y_2)}{1-x_2}\right) \ \}\, ,\nonumber
\end{eqnarray}
The arguments of the F1 function respect the absolute convergence criterium only on a portion of the range of the momentum. There are several functional relationships for the F1 function, but none of these capture the entire range of the momentum either. So far we have not found a single formula which works for all values of the momentum. For this reason we resort to numerical evaluation of the integral in the amplitude.

\section*{Acknowledgements}
We would like to thank Cosmin Crucean for guidance during the realization of this paper and Professor Ion Cotaescu for reading the manuscript and helpful remarks and discussions. \\
This work was supported by the strategic grant POSDRU/159/1.5/S/137750, Project ”Doctoral and Postdoctoral programs support for increased competitiveness in Exact Sciences research” cofinanced by European Social Fund within the Sectoral Operational Programme Human Resources Development 2007-2013.
\bibliographystyle{ws-mpla}
\bibliography{byblos}
\end{document}